\documentstyle[twocolumn,aps,epsfig,prl]{revtex}
\draft
\newcommand{\size}{5.7}

\begin{document}
\title{Internal localized eigenmodes on spin discrete breathers in
antiferromagnetic chains with on-site easy axis anisotropy}
\author{Sang Wook Kim$^1$ and Seunghwan Kim$^2$}
\address{$^1$Max-Planck-Institut f{\"u}r Physik komplexer Systeme,
N{\"o}thnitzer Stra{\ss}e 38, D-01187 Dresden, Germany\\
$^2$Asia Pacific Center for Theoretical Physics\\
Nonlinear and Complex Systems Laboratory, Department of Physics,\\
Pohang University of Science and Technology,
Pohang, Korea 790-784}
\date{\today}

\maketitle

\begin{abstract}
We investigate internal localized eigenmodes of the linearized equation 
around spin discrete breathers in 1D antiferromagnets with on-site easy 
axis anisotropy. The threshold of occurrence of the internal localized 
eigenmodes has a typical structure in parameter space depending on the 
frequency of the spin discrete breather. We also performed molecular 
dynamics simulation in order to show the validity of our linear analysis.
\end{abstract}


Discrete breathers (DB's) are the time-periodic and spatially
localized excitations on the nonlinear lattice with translational
invariance \cite{Flach98,Aubry97}, which are also called intrinsic
localized modes in the literatures \cite{Sievers88}. 
The existence of DB's was
mathematically proved in the anticontinuous limit \cite{Mackay91}.
Both the discreteness and the nonlinearity play a crucial role in
the existence of DB's. The spatial discreteness is quite common in
nature, particularly in condensed matter physics. Recently the
experimental realization of a DB was performed in anisotropic
antiferromagnets \cite{Schwarz99} and Josephson junction ladders
\cite{Binder00}.

In magnetic systems, both the spin-spin exchange interaction and
the on-site spin anisotropy are intrinsically nonlinear, so that
it is quite natural to predict the existence of DB's. Since the
dissipation of spin waves in magnetic materials is usually weak
compared with that of lattice vibrations in crystals, the spin
lattice model has obvious advantages over lattice vibration models
from the experimental point of views. Lai and Sievers have
extensively studied the DB's of spin wave, namely spin DB's,
for various situations of antiferromagnets \cite{Lai99}. The spin 
DB's have also been recently studied in ferromagnetic lattices
\cite{Zolotaryuk01}.

The DB can play a role of the scattering center affecting energy
transport by scattering, absorbing or radiating phonons. The scattering 
properties of the DB are closely related to the structure
of the eigenmodes of the linearized equation around the DB {\em itself}
\cite{Kim97,Kim00,Lee00,Kim01}, which will be called {\em internal}
localized eigenmodes (ILE's).
In particular, it was shown that the perfect transmission occurs 
at the ILE threshold, which should be tangent to the phonon
band edge with the zero wave number \cite{Kim00,Lee00}. However,
when the ILE's on DB's penetrate the phonon
band, the well known Fano resonances are obtained \cite{Kim01}.
These results can be directly applied to the spin lattice model
considering the analogy between the lattice vibration (phonon) and
the rotation of spin (magnon or spin wave). 

In this paper, we present the existence of the ILE's
on spin DB's using the linearized equation of spin DB's in one-dimensional
(1D) antiferromagnets with on-site easy axis anisotropy. The thresholds of
the ILE's show some special structure in parameter space.
The comparison between the prediction from linear analyses and the results
of molecular dynamics (MD) simulation will also be presented.

Let us consider the antiferromagnetic chain of N spins with the Hamiltonian
\cite{Lai96}
\begin{equation}
\label{hamiltonian}
H=2J\sum_n {\bf S}_n \cdot {\bf S}_{n+1} - D\sum_n (S_n^z)^2
,\end{equation}
where the positive $J$ and $D$ are the exchange constant and the single ion
anisotropy constant, respectively. Hence, the antiferromagnetic ordering
and aligning with $z$ direction of each spin are energetically more favorable
in the ground state.

Using the well known Heisenberg equation of motion
\begin{equation}
i\hbar \frac{d{\bf S}_n}{dt} = [{\bf S}_n,H]
\end{equation}
with $s_n^{\pm} = (S_n^x \pm iS_n^y)/S$, the nonlinear equation of motion
for $s_n^+$ can be obtained in the following
\begin{eqnarray}
\label{eq. of motion}
i\hbar \frac{ds_n^+}{dt} & = & -2JS[(s_{n-1}^z + s_{n+1}^z)s_n^+ - 
(s_{n-1}^+ + s_{n+1}^+)s_n^z] \nonumber \\
& + & 2DSs_n^z s_n^+
.\end{eqnarray}

If we assume a solution of the form $s_n^+ = s_n e^{-i\omega t}$, the stationary
spin DB can be obtained by using the following equation
\begin{eqnarray}
\label{db eq.} \tilde{\hbar}\omega s_n & = & (-1)^n
\{C(\sqrt{1-s_{n-1}^2}+\sqrt{1-s_{n+1}^2})s_n \nonumber \\
& + &[C(s_{n-1} + s_{n+1}) + s_n]\sqrt{1-s_n^2}\} 
,\end{eqnarray} 
where $\tilde{\hbar}=\hbar/2DS$ and $C=J/D$. In this paper we consider
only a single non-moving spin DB. The familiar dispersion relation
of the extended spin wave modes,
$\tilde{\hbar}\omega=\pm\sqrt{(1+2C)^2 - 4C^2 \cos^2ka}$, can also
be obtained by Eq. (\ref{db eq.}) by putting
\begin{equation}
\left(\begin{array}{c}
s_{2n}\\ s_{2n+1}
\end{array} \right)
=
\left(\begin{array}{c}
f_{0}e^{2ikna}\\ f_{1}e^{ik(2n+1)a}
\end{array} \right)
,\end{equation}
where $a$ is the lattice spacing, and $|f_{0}|$, $|f_{1}| \ll 1$.
The linearized equation of Eq. (\ref{eq. of motion}) near the spin DB for
$\xi_n (t) = s_n^+ (t) -s_n e^{-i\omega_b t}$ is given by
\begin{eqnarray}
\label{linearized eq.}
i\tilde{\hbar}\dot{\xi_n} & = & (-1)^n \{
C [ (A_{n-1}+A_{n+1})\xi_n - s_n(B_{n-1}Re(\xi_{n-1}) \nonumber \\
& + & B_{n+1}Re(\xi_{n+1})) + A_n(\xi_{n-1}+\xi_{n+1}) - B_n(s_{n-1} \\
& + & s_{n+1})Re(\xi_n) ] \nonumber + A_n\xi_n -s_n B_n Re(\xi_n) \} 
- \tilde{\hbar}\omega_b\xi_n, \nonumber
\end{eqnarray}
where $A_n=\sqrt{1-s_n^2}$, $B_n=s_n/2\sqrt{1-s_n^2}$, and $Re(\xi)$ is the
real part of $\xi$. The above Floquet (or linear)
equation is written in the form $d\vec{\xi}/dt=M\vec{\xi}$. By
diagonalizing the matrix $M$, the eigenmodes for small perturbation of
spin DB's are obtained. Let us note that there exist two important system
parameters, the relative strength of the exchange interaction, $C (=J/D)$
and the frequency of spin DB, $\omega_b$.

\begin{figure}
\includegraphics[height=\size cm]{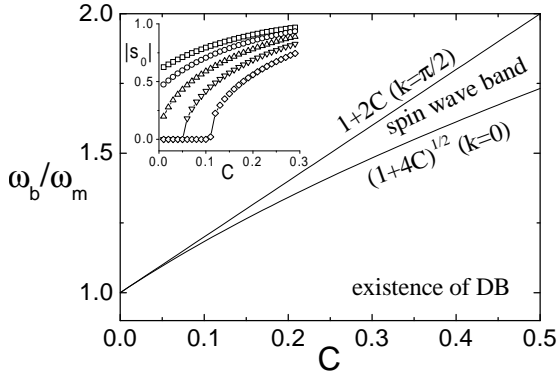}
\caption{
The band structure of spin waves as a function of $C$.
The inset represents the amplitude of the spin oscillation at the central site
of spin DB as a function of $C$ for a given frequency of spin DB $\omega_b$.
From the upper they correspond to $\omega_b/\omega_m =$ 0.8, 0.9
1.0, 1.1 and 1.2, respectively.}
\label{fig. band}
\end{figure}

Without the exchange interaction, i.e. $J=0$, the resonance
frequency of a local nonlinear oscillator is limited to
$\omega_m=1/\tilde{\hbar}$ since the nonlinearity of the
individual system is soft. In the case $\omega_b/\omega_m > 1$,
spin DB's cannot exist in the small coupling limit ($J \rightarrow
0$), so that we obtain two distinct regions for $\omega_b$ such as
$\omega_b/\omega_m> 1$ and $\omega_b/\omega_m < 1$. In the former
case ($\omega_b/\omega_m > 1$) the spin DB's can exist only above
the certain critical value of $C$, which can be explained by
considering a new on-site potential modified by the neighboring
sites. For simplicity let's consider just three spins, namely,
ones on the sites $-1$, $0$, and $1$, where the site $0$
corresponds to the center of the spin DB. In the weak coupling
limit the dynamics of the neighboring spins at the sites $\pm 1$
can be described by simple harmonic oscillation with a small
amplitude $\epsilon$, assuming the spin DB solution. From
Eq.~(\ref{db eq.}) we can obtain the following equation for the
frequency of the spin precession at the site $0$
\begin{equation}
\tilde{\hbar}\omega s_0 = 2Cs_0\sqrt{1-\epsilon^2}
+ (2C\epsilon + s_0)\sqrt{1-s_0^2}.
\end{equation}
Considering $\epsilon << s_0 \leq 1$ in the weak coupling limit,
the above equation can be approximated to $\omega/\omega_m \approx
2C + \sqrt{1-s_0^2}$, so that the new maximum frequency of local
nonlinear oscillators, namely $\omega_m'$, is given by
$\omega_m(2C+1)$, which means that the possible maximum frequency
of a spin DB also increases under the coupling with the
neighboring spins. Actually this relation exactly coincides with
the $k=\pi/2$ band edge of the spin wave as shown in
Fig.~\ref{fig. band} in the weak coupling limit. Taking into
account the instability caused by the resonance between DB's and
phonons \cite{Flach98}, we should exclude the overlapping region
between spin DB and the spin wave band, so that finally the border
of the existence of spin DB's is given by the $k=0$ band edge,
$\omega_m'/\omega_m=\sqrt{1+4C}$. The inset of Fig.~\ref{fig.
band} shows the amplitude of the spin oscillation at the central
site of spin DB as a function of $C$ for a given frequency of spin
DB $\omega_b$, where the criteria mentioned above can be also
confirmed numerically. In other words, the frequencies of spin
DB's are always located below the spin wave band, which is nothing
new in the sense of general criteria for the existence of a DB. We
would like to mention, however, that this leads to somewhat
interesting consequences that a spin DB appears above some
critical coupling strength, and disappears in the anticontinuous
limit with the frequency of a spin DB fixed. The mathematical
proof for the existence of DB's was performed in this limit
\cite{Mackay91}.

\begin{figure}
\includegraphics[height=\size cm]{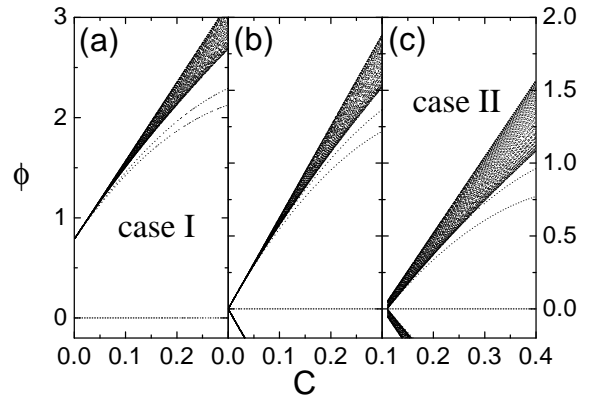}
\caption{Floquet spectrum for (a) $\omega_b/\omega_m=0.8$
(b) $\omega_b/\omega_m=1.0$, and (c) $\omega_b/\omega_m=1.2$.
See the left tick labels for (a) and the right tick labels for
both (b) and (c).}
\label{fig. floquet}
\end{figure}

Figure \ref{fig. floquet} shows the evolution of the eigenvalues of the Floquet
matrix $M$ on unit circle (the angle $\phi$ of the eigenvalue $e^{i\phi}$)
as a function
of $C$ for several $\omega_b$'s, where both spin wave bands and the detached
branches are clearly seen. For all calculations we take $\tilde{\hbar}=1$.
The eigenmodes coming from the bottom of the band are observed to be the ones
which are localized. These are ILE's mentioned above. The ILE's belong to either 
symmetric or antisymmetric eigenmodes with respect to their reflection symmetry to 
the center of a spin DB. While for $\omega_b/\omega_m < 1$ two ILE's appear
simultaneously at $C=0$ [Fig.\ref{fig. floquet}(a)], for $\omega_b/\omega_m > 1$
at first the symmetric ILE appears at a certain value of $C$ and
then the antisymmetric one appears at larger $C$ [Fig.~\ref{fig. floquet}(c)].
We call the former the case I and the latter the case II.

\begin{figure}
\includegraphics[height=\size cm]{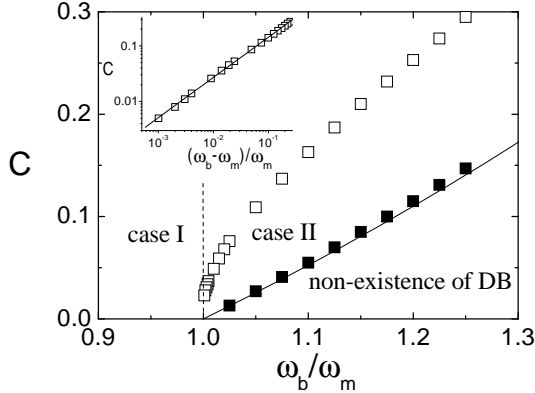}
\caption{The threshold of occurrences of symmetric (the filled squares) and
antisymmetric (the open squares) ILE's as a function of the frequency
of the spin DB. Spin DB's cannot exist in the region under the solid line
(see the text for details). In the inset we replot the threshold of the
antisymmetric ILE's (the open squares) in log-log scale, which
clearly shows power law dependence.}
\label{fig. threshold}
\end{figure}

Figure \ref{fig. threshold} shows the threshold of the ILE's as a function of 
$\omega_b$, where the dashed vertical line $\omega_b/\omega_m=1$, at which two 
spin wave bands collide each other as shown in Fig.~\ref{fig. floquet}(b),
divides the parameter space into two regions for the thresholds of
ILE's (i.e. the case I and II). The solid line represents the border
of the existence of spin DB's described by $C=[(\omega_b/\omega_m)^2-1]/4$,
below which the spin DB's cannot exist. It is mentioned that
the threshold of the symmetric ILE (the filled squares)
looks like the straight line, and that of the antisymmetric one
(the open squares) shows the power law dependence, namely
$C \propto \omega_b^{0.72}$ as shown in the inset of Fig.~\ref{fig. threshold}.
This threshold behavior was also observed in the case of Klein-Gordon (KG)
chain with $\phi^4$ (double well) on-site potential described by
$V(x)=(x^4-2x^2)/4$ \cite{skim}, where $\omega_b=2/3$, at which two neighboring
phonon band edges with $k=0$ collide each other, play a similar role
as $\omega_b/\omega_m=1$ in our antiferromagnet. However, this frequency does
not correspond to the maximum frequency of the local soft nonlinear oscillator,
which is $\omega_m=1$ in a $\phi^4$ KG chain. It should also be noted that
in other KG chains with Morse ($V(x)=[1-\exp(-x)]^2/2$) or cubic
[$V(x)=x^2/2-x^3/3$] on-site potential this is not the case, where for all
$\omega_b (<\omega_m = 1)$ the cubic and the Morse potential correspond only
to the case I and the case II, respectively. This has not been understood yet.

\begin{figure}
\includegraphics[height=\size cm]{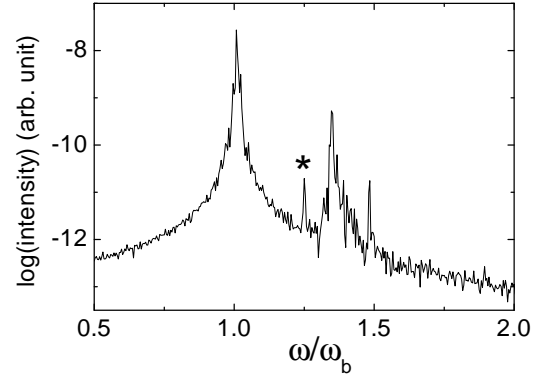}
\caption{The power spectrum of $M^y(t)$ with $\omega_b/\omega_m=1.1$ and $C=0.3$.
This is calculated using MD simulation with 103 spins during 256$T_b$
($T_b=2\pi/\omega_b$) under the random noise with the intensity $D=10^{-5}$ [14].
``$\star$'' indicates the symmetric ILE,
above which the spin wave excitations are shown.}
\label{fig. spectrum}
\end{figure}

One of the reasons why we choose this magnetic lattice for our
investigation is that experimentally the generation of a spin DB
was reported in the quasi-1D biaxial antiferromagnet
(C$_2$H$_5$NH$_3$)$_2$CuCl$_4$ \cite{Schwarz99} like the system
studied in this paper. By microwave absorption experiment the peak
corresponding to the spin DB was observed in the spin wave gap.
The absorption spectrum measured in the experiment is proportional
to the imaginary part of the dynamic magnetic susceptibility,
which can be calculated using the Kubo expression
\cite{Schwarz99}.
\begin{equation}
A(\omega) \propto \int_0^{\infty}dt
\left<M^y(t+t')M^y(t')\right>_{t'}e^{i\omega t} ,\end{equation}
where $M^y(t)=\sum_n s_n^y(t)$ and $\left<A\right>_{t'}$ is the
time average of the variable $A$. We calculate $M^y(t)$ using MD
simulations with 103 spins starting from the solution of a spin DB
obtained from Eq. (\ref{db eq.}) under the random amplitude noise
\cite{noise}. Figure \ref{fig. spectrum} clearly shows that an
absorption peak, which is a several order of magnitude smaller
than the spin DB peak, exists between the spin DB and the spin
wave band. Without noise we can obtain only the spin DB peak.
Although the absorption power of this ILE is so tiny
compared with that of the spin DB, it should not be ignored in
comparison with that of the spin wave band. It is worth noting
that only the symmetric ILE can be observed in the
absorption spectrum since the antisymmetric ILE disappears when
the summation, $\sum_n s_n^y(t)$, is performed. Figure \ref{fig.
comparison} shows that the frequency of the symmetric ILE
calculated using MD simulation is well fitted by the
prediction of the linear theory in the region of small $C$. It is
remarked that in the experiment of Ref.\cite{Schwarz99}, somewhat
broad spectrum of a spin DB was observed since many spin DB's with
different frequencies were excited simultaneously. In order to
observe the ILE's studied in this paper, more
improved experimental status will be needed, for example, a single
spin DB excitation, the suppression of volume and surface modes,
better resolution in the absorption power, and so on. We also note
that in general the antiferromagnetic ordering can persist only
above some critical value of $C$, roughly speaking the order of
one. Even though it does not seem to be easy to find out the
transition from the case I to the case II in this system in
experiment due to this reason, it is still possible to observe the
ILE's themselves for higher values of $C$.

\begin{figure}
\includegraphics[height=\size cm]{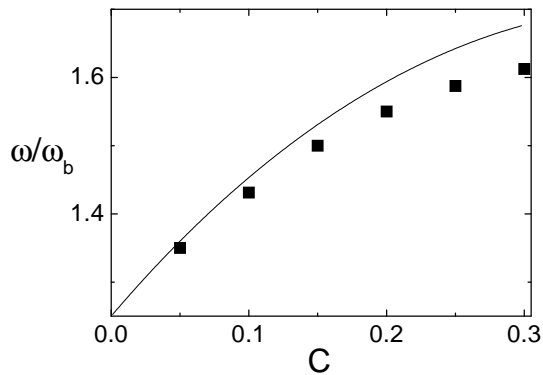}
\caption{The frequency of the symmetric ILE given by
the Floquet spectrum in Fig.~\ref{fig. floquet} (the solid curve)
and by the power spectrum of $M^y(t)$ using MD simulation (the filled square).}
\label{fig. comparison}
\end{figure}

In our previous work on discrete nonlinear Schr\"{o}dinger (DNLS)
equation \cite{Kim00} we already noted that the DNLS equation is
the nontrivial simplest system for studying scattering problems or
ILE's. However, it has only a single parameter
since the change of the breather frequency $\omega_b$ can be
compensated by scaling $\Omega \rightarrow \Omega/\omega_b$ and $C
\rightarrow C/\omega_b$, where $\Omega$ and $C$ are the frequency
of the small perturbation and the linear coupling strength,
respectively. We would like to point out that the 1D
antiferromagnet with easy axis anisotropy provides a good
intermediate example with two spin wave bands (or scattering
channels) and two parameters ($C$ and $\omega_b$) between the case
of DNLS equation with two phonon bands and one parameter
($C/\omega_b$) and that of the KG chain with an infinite number of
phonon bands and two parameters.

In summary we have investigated the characteristics and the
possibility of experimental observation of the ILE's
in 1D antiferromagnets with easy axis anisotropy using both
linear analyses and MD simulations. The structure of the
thresholds of the ILE's strongly depends on the
frequency of a spin DB. For $\omega_b/\omega_m < 1$ the ILE's
appear at $C=0$ simultaneously, while for $\omega_b/\omega_m
> 1$ the symmetric ILE occurs at first at the certain
value of $C$ and then the antisymmetric one appears at the larger
$C$. It is shown that the frequency of the ILE's
calculated using MD simulation is well fitted by the prediction of
the linear theory. We hope that these ILE's
will be observed in experiments.

%\section*{Acknowledgments}
We would like to thank Sergej Flach for a careful reading of this manuscript
and helpful comments. SW also thanks Mikhail Fistul for helpful discussions.

%%%%%%%%%%%%%%%%%%%%%%%%%%%%%%%%%%%%%%%%%%%%%%%%%%%%%%%%
%%%%%%%%%%%%%%%%%%%%%%%%%%%%%%%%%%%%%%%%%%%%%%%%%%%%%%%%

\bibliographystyle{prsty}

\begin{thebibliography}{10}
\bibitem{noise}
At each time step of numerical integrations of Eq.~(\ref{eq. of motion})
we add random amplitude noises $D\xi_n$ to the variables $s_n$, where the
absolute values of $\xi_n$ are normalized to one and $D$ represents the
strength of noise. $T_b/1024$ ($T_b=2\pi/\omega_b$) is used as the time
interval for numerical integration.
\end{thebibliography}

\end{document}